# Time Response Dynamics of Plasmon Excitation in Cobalt Nanoparticles on Glass Substrate

**Rajendra K. Shrestha[1] and Hernando Garcia[2]**

[1]Department of Physics. Oklahoma State University Univ cvgt. OM 9629:
[2]Department of Physics. Southern Illinois University Edwardsville. KN 84284

**Abstract**: We use ultrafast pump-probe transmission spectroscopy to measure the electronic relaxation time for electrons in cobalt nanoparticles embedded on glass substrate using femtosecond laser pulses. We found that the plasmon excitation is inhomogenously broadening with a fast excitation time of 23 fs. and a plasmonic relaxation time of 97fs. The femtosecond laser pulses used in this experiment were come from a Ti:Sapphire resonator that uses a self-mode-locking mechanism based on the nonlinear Kerr effect. The pulses were found to have repetition rate of 103.52MHz with a pulse width of 64.5 fs at FWHM, and 810nm center wavelength.

## References and links

## 10 Introduction and Theory

Electromagnetic radiation interacts with metals are largely dictated by the free conduction electrons in metals. According to simple Drude model [1] free electrons oscillate at $180^o$ out of phase relative to the applied electric field. As a consequence most metal possess a negative dielectric constant at optical frequencies which causes a high reflectivity. Furthermore at optical frequencies the metal's free electron gas can sustain at surface volume density oscillations called plasmon-polaritons or plasmons with distinct resonance frequencies. The existence of plasmons is the characteristic of the interaction of metal nanostructures with electric field. A plasmon is basically, an oscillation of conduction electrons in metal. It plays very important role in the optical properties of metals, for example: light of frequency lower than the plasma frequency gets reflected where as that with higher frequencies get transmitted.

The plasmons confined at the interface of the material having positive dielectric constant (glass) with that having negative dielectric constant (metal) is termed as surface plasmons (SPs), which has strong interaction with incident light resulting in a polaritons and gives rise to strongly enhanced optical near fields those are spatially confined near the metal surface. The charge density oscillations confined to metallic nanoparticles (metal clusters) and metallic nanostructures referred as localized surface plasmons (LSPs) results in strong light scattering, in appearance of intense surface Plasmon absorption bands.[3] Also, the frequency and intensity of the SP absorption bands which are the characteristic of the types of materials (gold, silver, copper...) are highly dependent to their size, size distribution, shape of the nanoparticles as well as the environment which surrounds them[2]

The metal nanoparticles embedded on a dielectric matrix constitute a unique area to study temporal evolution of the strongly correlated electron system. This is unique because surface Plasmons which is not diffraction limited, mediates the energy transfer between the electron and lattice. The development of nanostructured materials is an active area of research. [3]-[11] Electronic relaxation dynamics in metal nanoparticles occur on the ultrafast time scales and have been studied using femtosecond laser spectroscopy. As the electron-phonon energy relaxation time of most of the metals has been found to be in the order of few picoseconds, [6,8] the non-equilibrium between the electron and phonons can be achieved for intense optical pulses of duration shorter than or comparable to the electron-phonon relaxation time[8]. In case of electron gas, the specific heat capacity is small which allows the generation of an electron temperature much greater than the lattice temperature [11] and a transient inequality occurs. The evidence for non-equilibrium electron temperature has been reported from low intensity picoseconds reflectivity measurement in Cu. [10] However the time resolution was insufficient for the measurement of the electron-phonon energy transfer time.

In the case of metallic nanoparticles embedded in a dielectric matrix, the excitation and thermalization processes are strongly influenced by the scattering of electrons on the nanoparticle surface and the excitation of surface plasmon modes that mediate the energy transfer from electron to lattice subsystem.[5] The relaxation process is dominated by the electron-phonon interaction and the linear and nonlinear optical properties of the glass-metal nanocomposites show pronounced spectral features in the vicinity of the surface plasmon resonance (SPR). The size dependent electronic dynamics [11] showed that when the size of the metallic nanoparticle is smaller than the free electron mean free path, the relaxation process will be dominated by surface scattering. Many experiments regarding ultrafast optical measurements on metal nanoparticles both in solution and embedded in a matrix were performed in the past. [3]-[13]  For example, Mark J. Feldstein et al. [6] measured the electronic relaxation dynamics in thin films composed of 12-$nm$ colloid $Au$ nanoparticles. The hot electron lifetime were found to vary from 1 to 3 ps with the film's growth. Stella et al. [11] measured the size effect in the ultrafast electronic dynamics of metallic tin nanoparticles embedded in a matrix. Halonen et al. [5] measured the femtosecond absorption dynamics in glass-metal nanocomposites and so on.

If a metal is excited by a laser pulse with duration shorter than or comparable to the hot-electron energy loss life time ($\tau_e$), a transient inequality between the effective electron and lattice temperatures ($T_e$ and $T_l$) occurs. When an ultrafast optical pulse interacts with a metallic particle, the electrons absorb the energy first and thermalize rapidly through electron-electron (e-e) scattering and these electrons then transfer energy to the crystal lattice via electron-phonon (e-ph) coupling. Electrons thus excited to energy states of the metal above the Fermi level will relax initially by e-e scattering to the states near the Fermi level. For an incident laser pulse which is long compared to the e-ph energy transfer time, the electron and lattice will remain in thermal equilibrium but for a laser pulse comparable to or shorter than the e-ph energy transfer time, the electron and lattice will no more be in thermal equilibrium and the time evolution of electron and lattice will be given by coupled nonlinear differential equations [8]

$$c_e(T_e)\frac{\partial T_e}{\partial t} = K\nabla^2 T_e - g(T_e - T_l) + A(r,t) \qquad (1)$$

$$c_l\frac{\partial T_l}{\partial t} = g(T_e - T_l) \qquad (2)$$

Where $c_e$ is the electron heat capacity which is linearly dependent on the electron temperature, $c_l$ is the lattice heat capacity per unit volume, $K$ is the thermal conductivity and $A(r, t)$ represents the spatial and temporal profile of the heating source. g is the electron phonon coupling constant with value [8], $g \cong \dfrac{\pi^2 m n v_s^2}{6 \tau_{e-ph} T_l}$. Here $m$ and $n$ are respectively electron mass and number density, $v_s$ is the velocity of sound and $\tau_{e-ph}$ is the electron phonon collision time. These equations can be simplified to [3]

$$T_e(t) - T_l(t) = [T_e(0) - T_l(0)] \exp^{-t/\tau} \qquad (3)$$

Where „$\tau$" is the hot electron lifetime and is given by

$$\frac{1}{\tau} = \left( \frac{g}{c_e} + \frac{g}{c_l} \right) \qquad (4)$$

Here $T_e(0)$ and $T_l(0)$ are respectively the initial temperature of electron and lattice immediately following the excitation pulse. The temporal dependence of temperature difference, $T_e(t) - T_l(t)$ can be related to the experimentally measured transient reflectivity'[3]or differential transmission signal'[5,13].

## 2. Experimental Result and Discussion.

A Pump-probe experiment has been performed on a sample of cobalt nanoparticles on glass substrate. The experimental set up is shown in the Fig. 1. A stronger beam called the pump beam is used to excite the sample and a weaker beam called the probe beam carries the information from the excitation of the material with the pump beam. In our experiment, both the pump and probe beams are the femtosecond laser pulses. The laser pulses at wavelength of 810nm are produced by the self-mode-locking technique from the Ti:Sapphire laser cavity [15-19] based on nonlinear optical Kerr effect.[20] The pulse width at FWHM is 64.5 $fs$ and the repetition rate is 103.6 MHz with a secant hyperbolic squared profile as shown in Fig. 2.

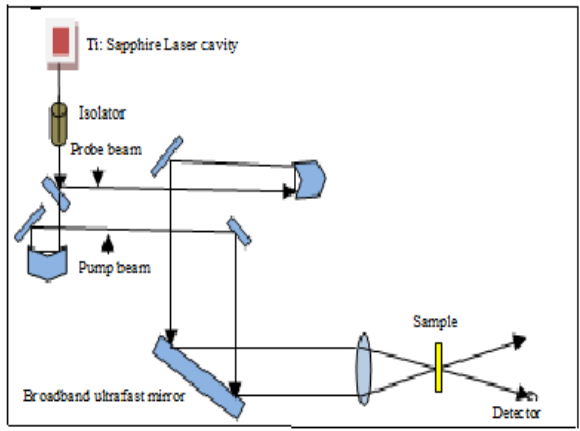

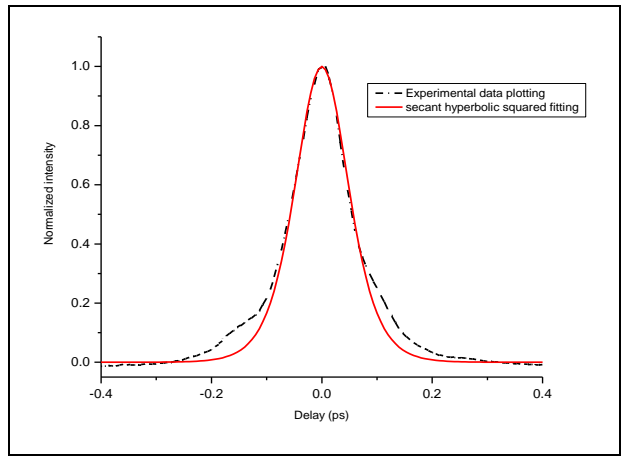

Fig.1. Experimental set up for pump-probe experiment.     Fig.2. Secant hyperbolic fitting to the femtosecond laser pulse

After passing through an isolator, the beam is split into two beams with 90/10 beam splitter and then both beams are focused on the sample. The pump beam at the sample has average power of 150 mW, the intensity at focus is $47 \times 10^6 \mathrm{W/m^2}$ and a fluence of 0.0912 mJ/cm². The two beams should have spatial and temporal overlap to get the information from the sample, which is very important in pump- probe experiment. The pump beam is made to have a variable time delay relative to probe beam with the help of automated computer programming. Both the beams are chopped with different frequencies and the signal is detected at sum frequencies using DET 110 high speed silicon detector and lock-in amplifier.

## 3. Result and Discussion

A pump- probe experiment has been performed on a sample of cobalt nanoparticles on glass substrate. The information about the system carried out by probe beam was detected as differential transmission versus delay time as shown in Fig.3.

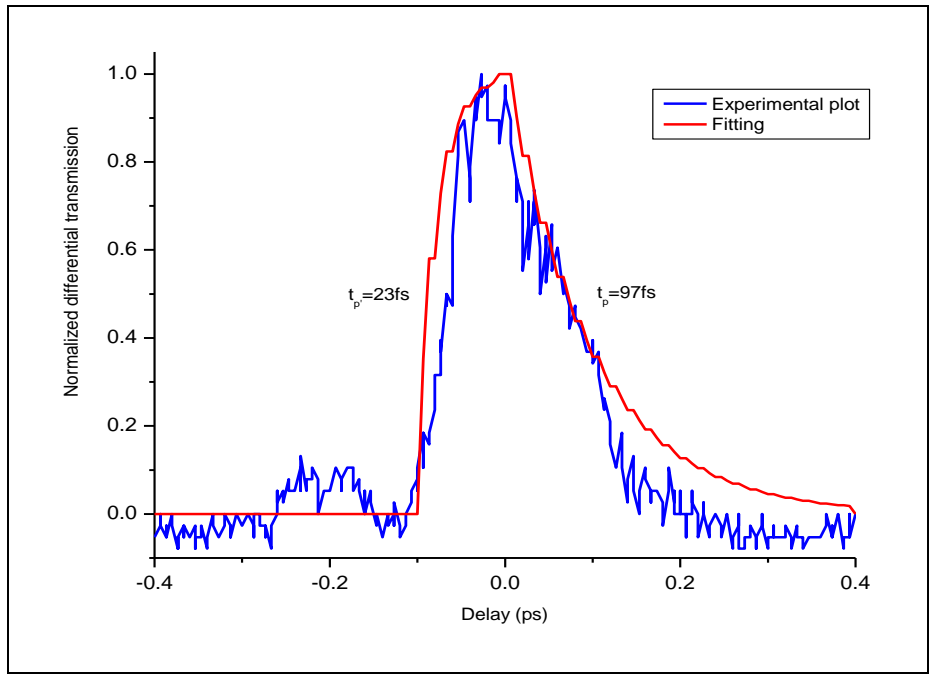

Fig.3. Plot of normalized transmission versus delay time with exponential fitting

The plot shows that at the zero delay, the transmission of the signal peaks and then starts to decay with increasing delay time. The fitting shows that the Plasmon excitation is inhomogeneously broadened with a fast excitation time of 23 fs. and a Plasmon relaxation time of 97fs. This result is close to the known values of electron relaxation time in cobalt which is in the order of 120fs [3] Theoretically, the inhomogeneous broadening should be ¼ of the relaxation time.[14]. The inhomogeneous broadening is due to the different line shape of the atoms in the cobalt nanoparticles.